\documentclass{ws-ijmpa}

\def\lsim{\stackrel{<}{{}_\sim}}
\def\gsim{\stackrel{>}{{}_\sim}}

\begin{document}

\markboth{Jae Sik Lee}
{LHC Signatures of MSSM Higgs-sector CP Violation}

%
\catchline{}{}{}{}{}
%

\title{LHC Signatures of MSSM Higgs-sector CP Violation }

\author{JAE SIK LEE}

\address{Center for Theoretical Physics, School of Physics and Astronomy,\\
Seoul National University, Seoul 151-747, Korea\\
jslee@muon.kaist.ac.kr}

\maketitle

\begin{history}
\received{15 June 2007}
\end{history}

\begin{abstract}
We discuss a few new characteristic features of the loop-induced
MSSM Higgs-sector CP violation at the LHC based on two scenarios:
$(i)$ CPX and $(ii)$ Trimixing.

\keywords{Higgs; CP violation; LHC.}
\end{abstract}

\ccode{PACS numbers: 14.80.Cp, 12.60.Jv, 11.30.Er}

\section{Introduction}
Supersymmetric models contain many possible sources of 
CP violation beyond the SM CKM phase. In
the Minimal Supersymmetric extension of the Standard Model (MSSM), for example,
we have 8 CP phases when we even consider only the third generation, that is,
stops, sbottoms, and staus:
\begin{itemize}
\item $\Phi_\mu\,[1]$: $~~~~W\,\supset\, \mu\, \hat{H}_2\cdot\hat{H}_1$
\vspace{0.2cm}
\item $\Phi_i\,[3]$: 
$~~
-{\cal L}_{\rm soft}\, \supset\, \frac{1}{2}
( M_3 \, \widetilde{g}\widetilde{g}
+ M_2 \, \widetilde{W}\widetilde{W}
+ M_1 \, \widetilde{B}\widetilde{B}+{\rm h.c.})$
\vspace{0.2cm}
\item $\Phi_{A_f}\,[3]$ with $f=t,b,\tau$:\\
$~~~~~~~~~~~~
-{\cal L}_{\rm soft}\, \supset\, 
 A_t \, \widetilde{t}_R^*\,\widetilde{Q}_3\cdot H_2
-A_b \, \widetilde{b}_R^*\,\widetilde{Q}_3\cdot H_1
-A_\tau \, \widetilde{\tau}_R^*\,\widetilde{L}_3\cdot H_1+{\rm h.c.}$
\vspace{0.2cm}
\item $\Phi_{m_{12}^2}\,[1]$:
$-{\cal L}_{\rm soft}\, \supset\, -(m_{12}^2\,H_1\cdot H_2 + {\rm h.c.})$
\end{itemize}
The numbers of relevant CP phases are given in the brackets.
These 8 CP phases
are not all independent and physical observables depend on the combinations
of ${\rm Arg}\left(M_i\mu (m_{12}^2)^*\right)$ and
${\rm Arg}\left(A_f\mu (m_{12}^2)^*\right)$.\cite{Dugan:1984qf,Dimopoulos:1995kn}
In the convention of ${\rm Arg}(m_{12}^2)=0$, we have 6 rephasing invariant CP phases:
\begin{equation}
{\rm Arg}(M_1\,\mu)\,, \ \
{\rm Arg}(M_2\,\mu)\,, \ \
{\rm Arg}(M_3\,\mu)\,; \ \
{\rm Arg}(A_t\,\mu)\, \ \
{\rm Arg}(A_b\,\mu)\, \ \
{\rm Arg}(A_\tau\,\mu)\,.
\end{equation}

These non-vanishing CP phases can induce a significant
CP-violating mixing between CP-even and CP-odd Higgs states
via radiative corrections.\cite{Pilaftsis:1998pe,Pilaftsis:1998dd,Pilaftsis:1999qt,Demir:1999hj,Choi:2000wz,Ibrahim:2000qj,Ibrahim:2002zk}
There are two approaches to calculate this CP-violating mixing.
Here we use the calculation based 
on the renormalization-group-improved effective potential
method including the Higgs-boson pole mass 
shift.\cite{Carena:2000yi,Carena:2001fw}
For the Feynman-diagrammatic approach, 
we refer to Ref.~\refcite{Heinemeyer:2007aq} and references there in.

In this contribution, we discuss a few characteristic features of the 
Higgs-sector CP violation at the LHC which have been recently
observed after the appearance of
{\bf CPNSH} Report.\cite{Accomando:2006ga}
And we put emphasis on importance of the 
$\tau$-lepton polarization measurement to construct genuine CP-odd signal at the LHC.
\cite{Ellis:2004fs,Ellis:2006eh}
For numerical analysis, two scenarios are considered:
$(i)$ CPX\cite{Carena:2000ks} (Sec. 2)
and $(ii)$ Trimixing\cite{Ellis:2004fs} (Sec. 3). 
See, for example, Ref.~\refcite{Lee:2007gz}
for detailed description and comparison of two scenarios
with some numerical results.
The code {\tt CPsuperH}\cite{Lee:2003nt} is used to generate
numerical outputs.

\section{CPX Scenario}

First we consider the constraint on the CPX scenario coming from the
non-observation of an EDM in the Thallium atom. 
The contributions  of
the   first  and second  generation  phases,  e.g.~$\Phi_{A_{e,\mu}}$,
$\Phi_{A_{d,s}}$ etc.,  to EDMs can be  drastically reduced either by
making these phases  sufficiently   small, or if the first- and
second-generation squarks and  sleptons are sufficiently heavy.
In this case, the dominant contribution to EDMs occurs at two-loop 
level.\cite{Chang:1998uc,Pilaftsis:1999td,Chang:1999zw}
We refer to Ref.~\refcite{Ellis:2005ik} for the explicit expression of the
two-loop Higgs-mediated Thallium EDM 
in the {\tt CPsuperH} conventions and notations.

\begin{figure}[htbh]
\begin{center}
\includegraphics[width=6.2cm]{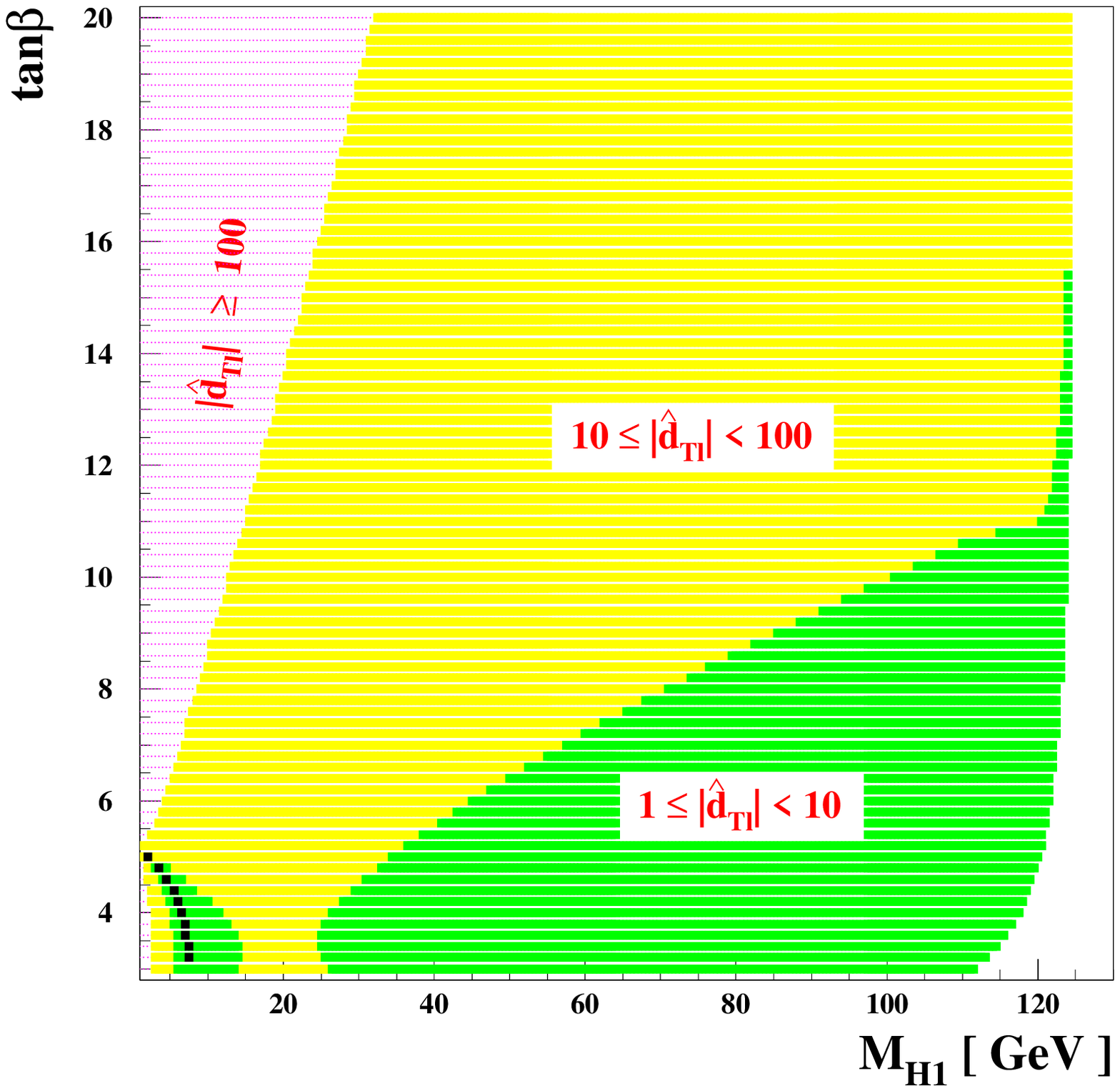}
\includegraphics[width=6.2cm]{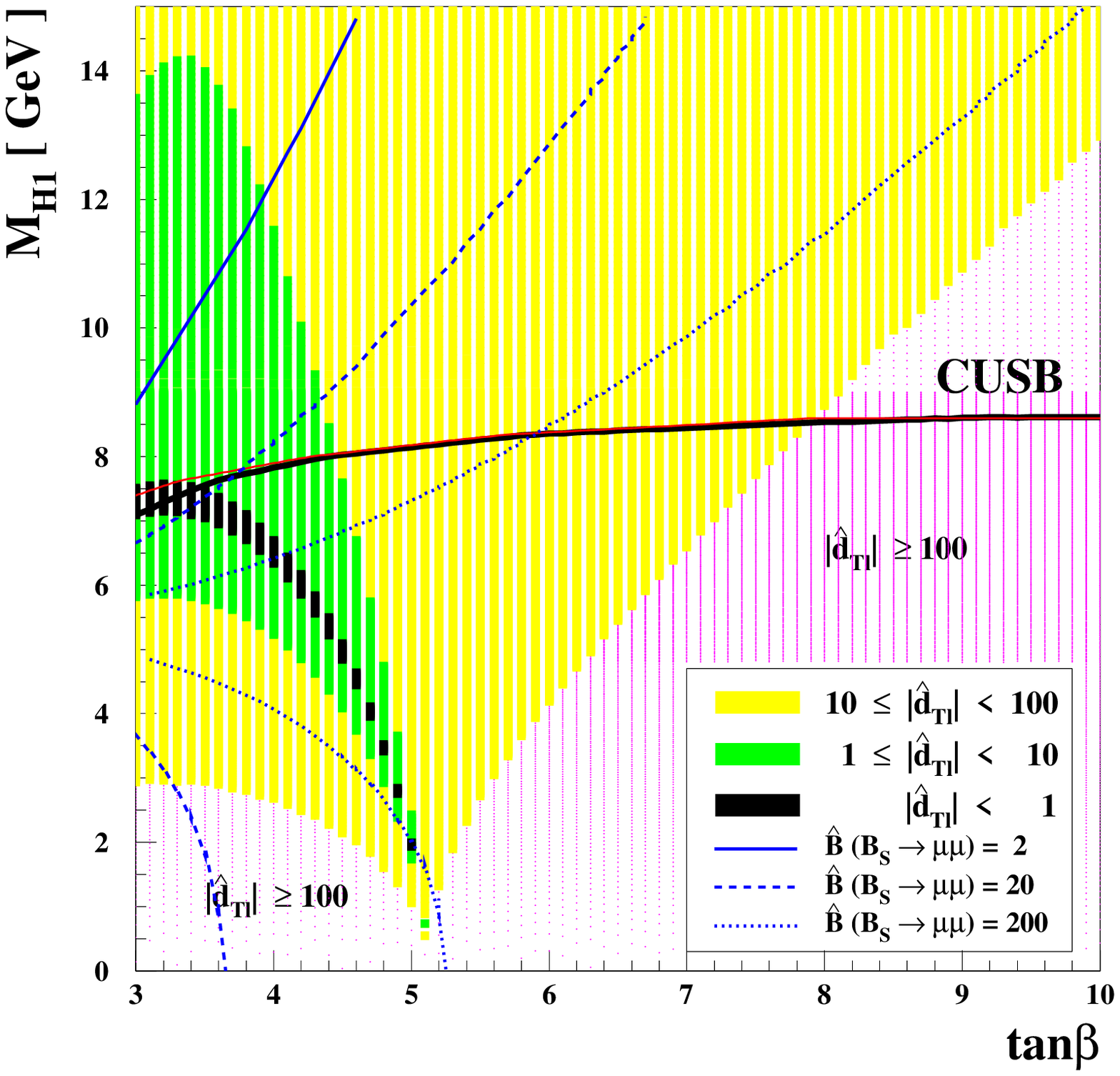}
\end{center}
\caption{The rescaled Thallium EDM 
$\hat{d}_{\rm Tl}\equiv d_{\rm Tl}\times10^{24}$
in units of $e\,cm$ for the CPX scenario with 
$\Phi_{A_{t,b,\tau}}=\Phi_3=90^\circ$ on the
$\tan\beta-M_{H_1}$ plane. We take $\Phi_\mu=0$ convention.
In the right frame the CUSB bound from the decay
$\Upsilon(1S)\rightarrow \gamma H_1$ is also shown as a thick solid line. See
Ref.~\protect\refcite{Lee:2007ai} for details.}
\label{fig:dtl}
\end{figure}
In the left frame of Fig.~\ref{fig:dtl}, the
rescaled Thallium EDM $\hat{d}_{\rm Tl}\equiv d_{\rm Tl}\times10^{24}$ is shown on the
$\tan\beta-M_{H_1}$ plane in units of $e\,cm$. The current upper limit is 
$|\hat{d}_{\rm Tl}|\lsim 1.3$.\cite{Regan:2002ta} We divide the plane 
into 4 regions depending on the size of $|\hat{d}_{\rm Tl}|$. The 
unshaded region is not allowed theoretically. We have 
$|\hat{d}_{\rm Tl}|< 1$ only in the narrow region 
filled with black squares when $\tan\beta\lsim 5$ and $M_{H_1}\lsim 8$ GeV.
However, if we allow 10 \%-level cancellation between the two-loop contributions and possible
one-loop contributions not considered here, the (green) region with 
$1 \leq |\hat{d}_{\rm Tl}|< 10$ is allowed. Furthermore, 
if very strong 1 \%-level cancellation is possible, 
most of the region can be made consistent with the Thallium
EDM constraint if the lightest Higgs boson is not so light.
In the right frame of Fig.~\ref{fig:dtl}, we magnify the region with 
$3 \lsim \tan\beta \lsim 10$ and $M_{H_1} \lsim 15~{\rm GeV}$. This region
is of particular interest since $H_1$ lighter than  about 10 GeV has not been
excluded by the LEP experiments for the 
given rage of $\tan\beta$.\cite{Schael:2006cr,bechtle}  The bound on this light Higgs boson
comes from low-energy experiment. We find that the region
$M_{H_1}\lsim 8$ GeV (the region
below the thick solid CUSB line)  is excluded by data on $\Upsilon(1S)$
decay.\cite{Franzini:1987pv} For details, see Ref.~\refcite{Lee:2007ai}.
\begin{figure}[htbh]
\begin{center}
\includegraphics[width=6.2cm]{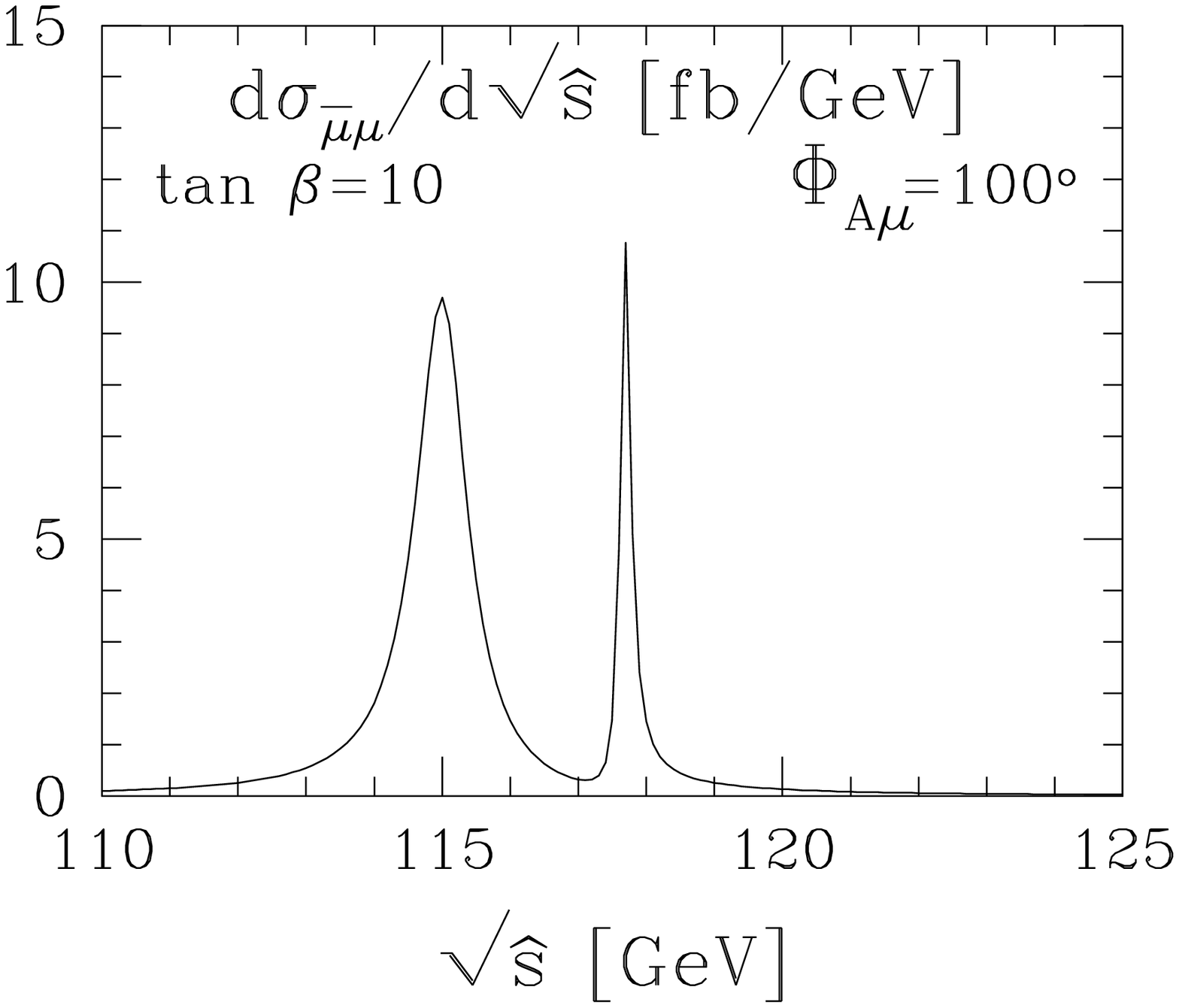}
\includegraphics[width=6.2cm]{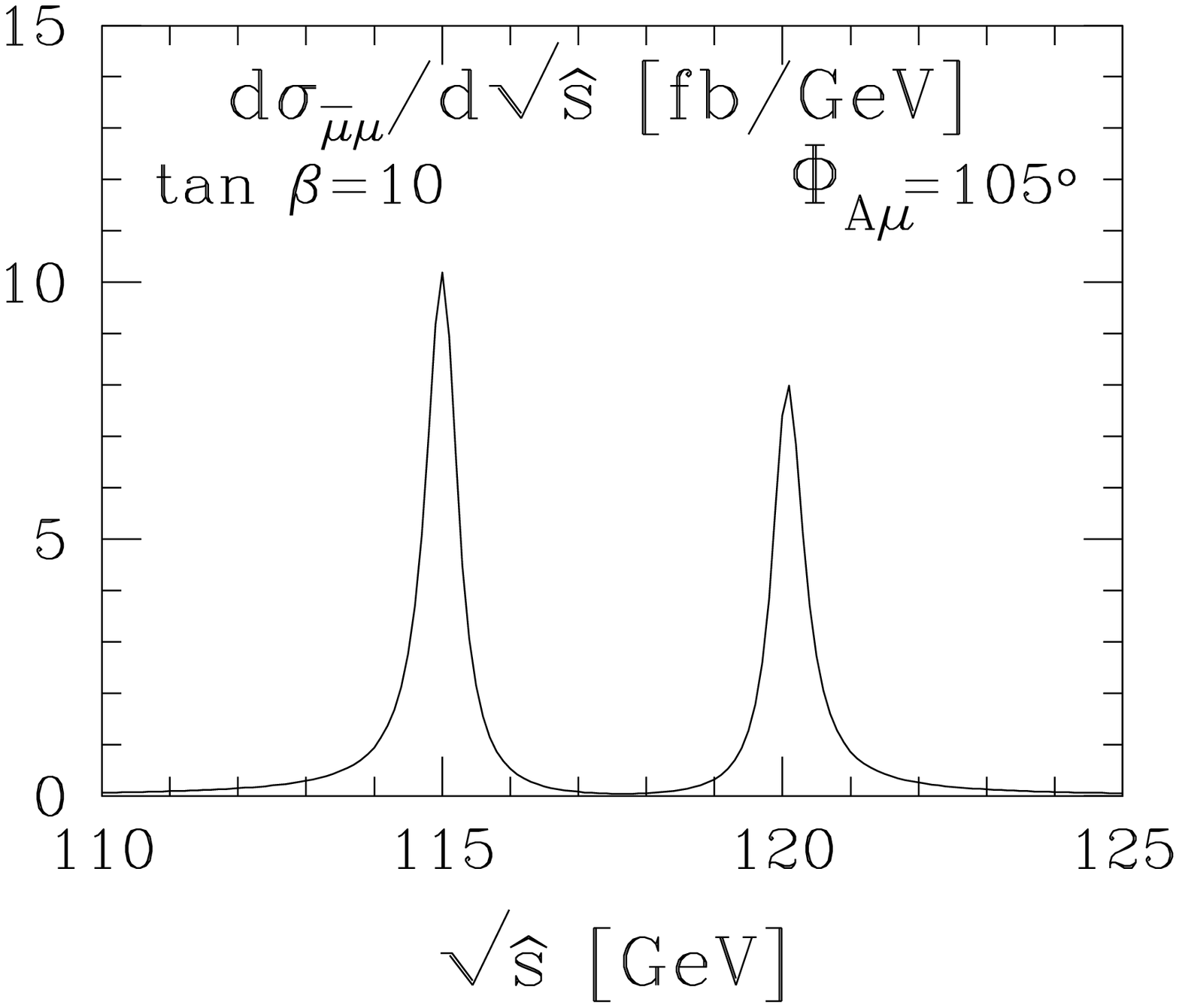}
\includegraphics[width=6.2cm]{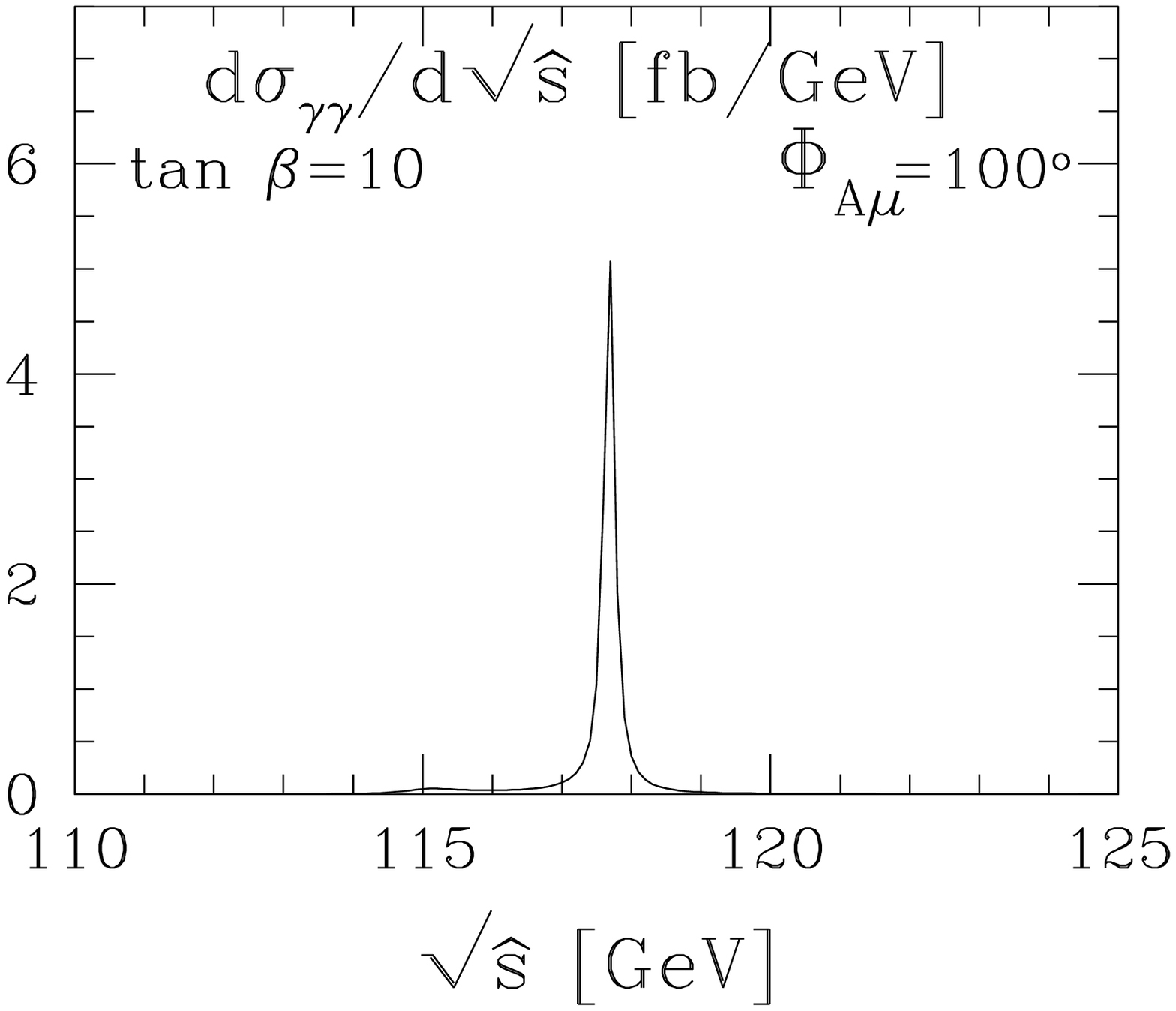}
\includegraphics[width=6.2cm]{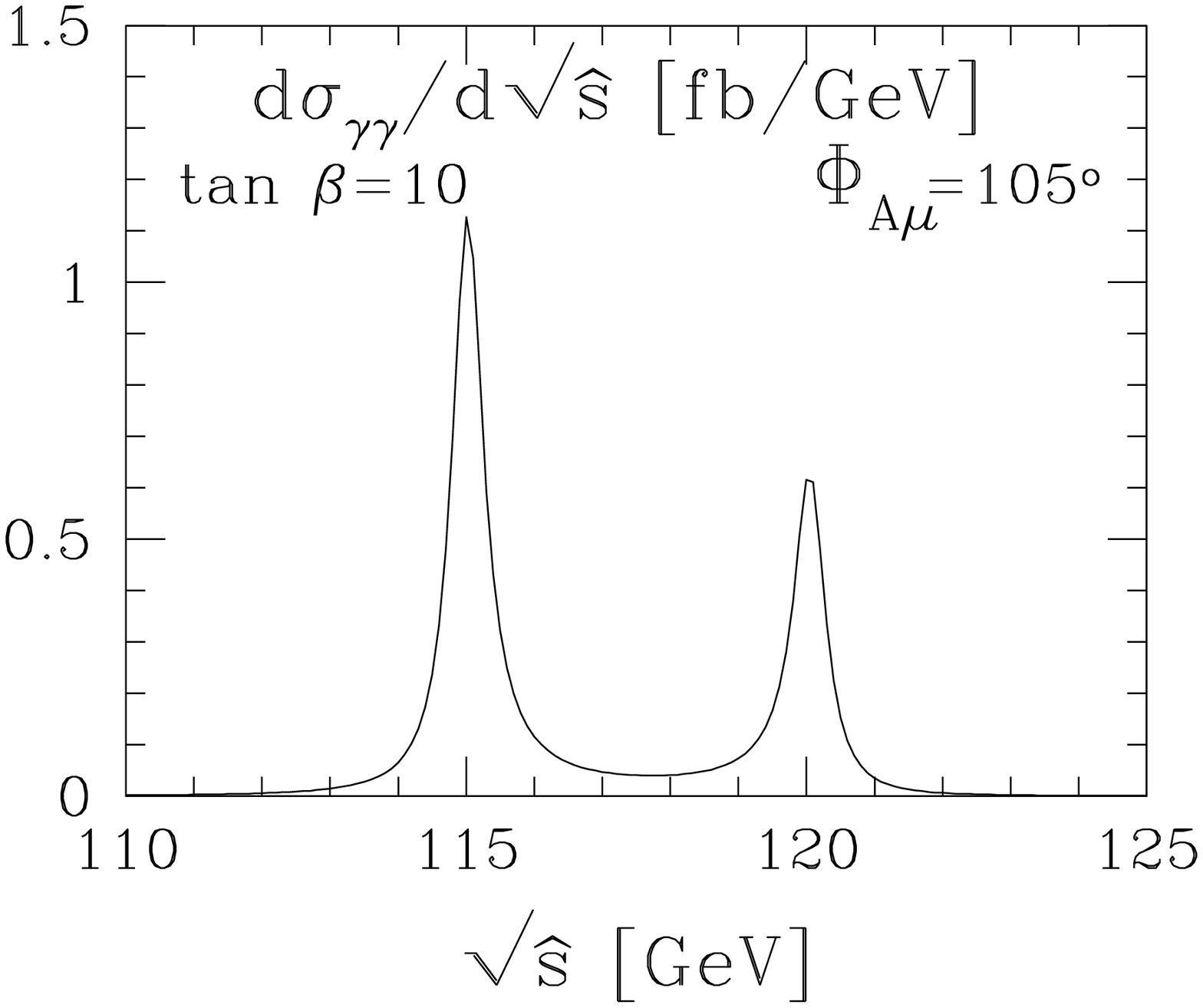}
\end{center}
\caption{The differential cross sections in units of fb/GeV at
$\Phi_{A\mu} = 100^{\circ}$ (left frames) and 
$\Phi_{A\mu} = 105^{\circ}$ (right frames),
versus the invariant mass $\sqrt{\hat{s}}$
of two muons (uppers frames) or two photons (lower frames).
The charged Higgs-boson pole mass is solved to give $M_{H_1}=115$ GeV
for $\tan \beta = 10$ and  ${\rm Arg}(M_3\,\mu) = 180^{\circ}$ in the CPX
scenario. See Ref.~\protect\refcite{Borzumati:2006zx} for details. }
\label{fig:diff}
\end{figure}
For the scenario with large $|\mu|$ and $|M_3|$ such as CPX, the threshold 
corrections
to the bottom-quark Yukawa coupling should not be neglected especially
for intermediate and large values of $\tan\beta$.
In this case, the production cross sections of the three neutral Higgs
bosons through $b\bar{b}$ fusion can deviate substantially from
those obtained in CP conserving scenarios, thanks to the
nontrivial role played by the threshold corrections
combined with the CP-violating mixing in the neutral-Higgs-boson
sector.\cite{Borzumati:2004rd}
The largest deviations in the case of $H_1$ and $H_2$
are for values of $\Phi_{A\mu}\equiv {\rm Arg}(A_{t,b}\,\mu)$
around $100^\circ$, with a large enhancement for
the production cross section of $H_1$ and a large suppression for that of $H_2$.
To detect this large  enhancement and/or suppression, we need to know
whether it is possible to disentangle the two corresponding peaks in the
invariant mass distributions of the $H_1$- and $H_2$-decay products
at the LHC. To address this issue, we consider 
the Higgs-boson decays into muon and photon pairs.
For these two decay modes, the invariant-mass resolutions
are, respectively,
$\delta M_{\gamma\gamma}\sim 1\,$GeV and
$\delta M_{\mu\mu} \sim 3\,$GeV for a Higgs mass of
$\sim 100\,$GeV.\cite{ATLASTDR:1999fr}
In Fig.~\ref{fig:diff}, we show the 
differential cross sections in units of fb/GeV taking two values of 
$\Phi_{A\mu}$.
The upper two frames are for $H_{1,2}\rightarrow \mu^+\mu^-$ and
the lower frames for $H_{1,2}\rightarrow \gamma\gamma$.
For $\Phi_{A\mu}=100^{\circ}$ (left frames),
by combining the muon-decay mode with the
photon-decay mode, $H_2$ can be located more precisely
and disentangled from $H_1$. For
$\Phi_{A\mu}=105^\circ$ (right frames), actually, two well separated
peaks may be observed. For details, we refer to Ref.~\refcite{Borzumati:2006zx}.

\section{Trimixing Scenario}

\begin{figure}[htbh]
\begin{center} \includegraphics[width=12cm]{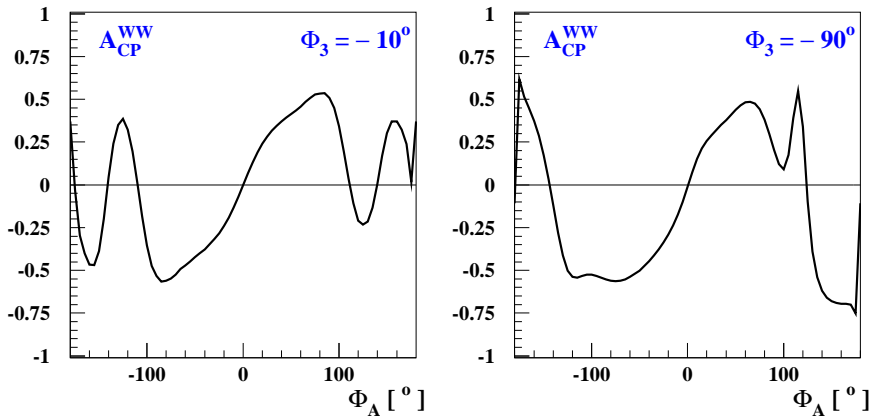}
\end{center} 
\caption{The CP asymmetry ${\cal A}^{WW}_{\rm CP}$ as functions of
$\Phi_A\equiv\Phi_{A_t}=\Phi_{A_b}=\Phi_{A_\tau}$ for $\Phi_3=-10^\circ$ (left
frame)
and $\Phi_3=-90^\circ$ (right frame) in the Trimixing scenario. We take
$\Phi_\mu=0$. 
See Ref.~\protect\refcite{Ellis:2004fs} for details. }
\label{fig:acpww} 
\end{figure}

\noindent
To construct CP asymmetry at the LHC, we consider the
production of CP-violating MSSM $H_{1,2,3}$  bosons via 
$W^+ W^-$ collisions and their subsequent decays into $\tau^+
\tau^-$  pairs assuming the longitudinal polarization of
$\tau$ leptons can be measured.\cite{Ellis:2004fs} In this case, 
one can define integrated CP asymmetry:
\begin{equation}
  \label{CPasym}
{\cal A}^{WW}_{\rm CP} \ \equiv \
\frac{ \sigma^{WW}_{\rm RR}-\sigma^{WW}_{\rm LL} }{\sigma^{WW}_{\rm RR}+\sigma^{WW}_{\rm LL}}\ ,
\end{equation}
where
\begin{eqnarray}
\sigma_{RR}\ &=&\ \sigma(pp (WW)\ \to\ H\ \to\  \tau^+_R\tau^-_R X) \,, \nonumber \\
\sigma_{LL}\ &=&\ \sigma(pp (WW)\ \to\ H\ \to\  \tau^+_L\tau^-_L X) \, .
\end{eqnarray}

In Fig.~\ref{fig:acpww}, 
we show the CP asymmetry ${\cal A}^{WW}_{\rm CP}$ as functions of
$\Phi_A\equiv\Phi_{A_t}=\Phi_{A_b}=\Phi_{A_\tau}$ for $\Phi_3=-10^\circ$ (left frame)
and $\Phi_3=-90^\circ$ (right frame) taking $\Phi_\mu=0^\circ$. 
We observe the CP asymmetry is large over the whole
region of $\Phi_A$ independently of $\Phi_3$. For more detailed discussion, 
see Ref.~\refcite{Ellis:2004fs}.

\section{Conclusions}
We obtain the constraint $M_{H_1}\gsim 8$ GeV from the decay
$\Upsilon(1S)\rightarrow \gamma H_1$.
By combining the Higgs-boson decay mode into two muons with that into
two photons,
it is possible to disentangle two adjacent peaks with the mass 
difference larger than $\sim 3$ GeV at the LHC.
The process $W^+ W^- \to  H_{1,2,3}  \to \tau^+  \tau^-$ is
promising  to probe CP violation through
the CP  asymmetry   based  on 
longitudinal  $\tau$-lepton  polarization.

\section*{Acknowledgments}

I wish to thank F. Borzumati, J. Ellis, S. Scopel, and A. Pilaftsis for valuable
collaborations. This work was supported in part by Korea Research Foundation and
the Korean Federation of Science and Technology Societies Grant
funded by the Korea Government (MOEHRD, Basic Research Promotion Fund).

\section{References}


\end{document}